\documentclass[aps,pra,twocolumn,longbibliography]{revtex4-1}
\usepackage{epsfig,amsopn}
\usepackage{graphicx}
\usepackage{color}
\usepackage{amsmath,amssymb}
\usepackage{enumerate}
\newcommand\bea{\begin{eqnarray}}
\newcommand\eea{\end{eqnarray}}
\newcommand\beq{\begin{equation}}
\newcommand\eeq{\end{equation}}

\def\nn{\nonumber}
\def\f{\frac}

\def\om{\omega}
\def\de{\delta}

\def\ga{\gamma}

\def\si{\sigma}

\def\De{\Delta}
\def\dg{\dagger}

\def\ra{\rangle}
\def\ua{\uparrow}
\def\da{\downarrow}

\def\th{\theta}

\begin{document}
\title{Nonequilibrium Josephson diode effect in periodically driven SNS junctions} 
\author{ Abhiram Soori~~}  
\email{abhirams@uohyd.ac.in}
\affiliation{ School of Physics, University of Hyderabad, C. R. Rao Road, 
Gachibowli, Hyderabad-500046, India.}
\begin{abstract}
In typical Josephson junctions, the Josephson current is an odd function of the superconducting phase difference. Recently, diode effect in Josephson junctions is observed in experiments wherein the maximum and the minimum values of the Josephson current in the current-phase relation  do not have the same magnitude. We propose a superconductor-normal metal-superconductor (SNS) junction where Josephson diode effect manifests when the  normal metal region  is driven. Time reversal symmetry and inversion symmetry need to be broken in the SNS junction for the diode effect to show up.  We calculate long time averaged current and show that the system exhibits diode effect for two configurations of the driven SNS junction - one in which inversion symmetry is broken in the undriven part of the Hamiltonian  and the other wherein both the symmetries are broken by the driving potential. In the latter configuration, a nonzero current known as anomalous current appears at the junction in absence of phase bias. In the proposed setup, the diode effect vanishes in the adiabatic limit.  
\end{abstract}
\maketitle

\section{Introduction}

The phenomenon of rectification, wherein the two terminal resistance across an electrical device depends on the direction of current, dates back to 1874~\cite{braun}. Semiconductor diodes that exhibit this phenomenon consist of a pn-junction and the reason behind rectification is rooted in the fact that the widths of the  depletion region in the forward and reverse bias are different~\cite{malvino}. However, such a non-reciprocal transport, also known as diode effect, is a purely classical effect in semiconductor diodes. When the transport is quantum coherent, the scattering matrix is unitary and the rectification  is not possible in a two terminal setup with normal metal leads~\cite{imry}. In scattering across interacting regions that break inversion symmetry, though the two-particle scatterings are direction dependent, it does not lead to rectification in current-voltage characteristics~\cite{roy09}. Decoherence in quantum devices can lead to rectification~\cite{bredol}. It may be noted that in certain devices, the four probe resistance exhibits non-reciprocal quantum coherent transport~\cite{song98,ideue17,legg22}. 

The current-voltage characteristics of a superconductor is accompanied by a critical current, below which the voltage drop is zero. In the context of superconductors, the magnitude of critical current through the superconductor being direction dependent marks diode effect. In recent years, the diode effect was observed in superconductors under an applied magnetic field~\cite{waka17,qin17,siva18,lust18,hoshino18,yasuda19,ando2020,ita2020,baum2022,moodera,suri22}.   Time reversal symmetry and inversion symmetry need to be broken to realize the diode effect in superconductors. While in some of these works~\cite{waka17,hoshino18,yasuda19,ando2020,ita2020,baum2022}, the magnetic field in combination with spin-orbit coupling gives rise to magnetochiral anisotropy, vortex dynamics is responsible for the observed diode effect in a few other works~\cite{lust18,siva18}. Across a junction between two superconductors differing by a phase, a current that depends on the phase difference flows -an effect known as Josephson effect~\cite{josephson62,anderson63,furusaki99}. The maximum and minimum values of the current as the phase difference is varied are known as critical currents in forward and backward directions.  In an inversion symmetry broken Josephson junction setup, the Josephson critical currents in the forward and backward directions are found to be different in magnitude without a need for magnetic field and the origin of the diode effect is debated~\cite{wu2022}. An interferometric setup has been proposed wherein a magnetic flux threaded between two channels that connect phase biased superconductors  can give rise to Josephson diode effect with efficiencies as large as 40\%~\cite{schrade}.

Periodically driven systems have been studied with a renewed interest in recent years, since novel phases of matter can be realized in such out of equilibrium systems~\cite{oka19,bandho21}. Tight binding model has been used to study a variety of transport phenomena in periodically driven systems~\cite{gomez13,Longhi2017,Mei2022}. Rectification and anomalous current in junctions between Floquet superconductors has also been investigated~\cite{soori23aje}. Quantum charge pumping is a phenomenon wherein a small region connected to normal metal leads is driven periodically, giving rise to a net flow of charge in the absence of a potential difference~\cite{thouless83,brouwer98,switkes99,brouwer01,
 avron01,moskalets02,agarwal07,agarwal07prb,soori10}. Quantum charge pumping across regions connected to superconducting leads is investigated  by some groups~\cite{blaauboer02,governale05,soori20pump}. In a phase biased superconductor-normal metal-superconductor~(SNS) junctions, driving the normal metal (NM) region can block the Josephson current and  transfer a net charge from one superconductor to another in the absence of a superconducting phase difference~\cite{soori20pump}. The net current averaged over a long time quantifies the current flow. We show that in such an SNS junction, by driving the NM region, the Josephson critical currents in the forward and backward directions can be made different in magnitude in two ways. In one way, the inversion symmetry is broken in the undriven Hamiltonian and the long time averaged current is zero when the phases of the two superconductors are the same. In the second way, the undriven system is inversion symmetric and the time dependent potentials in NM region break inversion and time reversal symmetries, wherein a nonzero long time averaged current appears at zero superconducting phase difference. 
 
 The manuscript is structured as follows. In sec.~\ref{sec-model}, the model is described with details of calculation. In sec.~\ref{sec-res}, the results are explained and analyzed. In sec.~\ref{sec-con}, we summarize the work and conclude. 

  \section{Model and calculation}~\label{sec-model}

\begin{figure}[htb]
 \includegraphics[width=9cm]{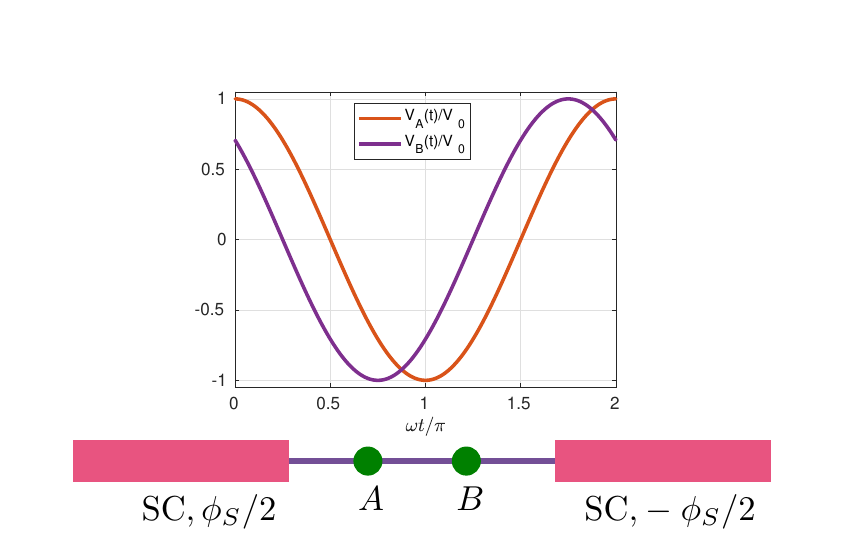}
 \caption{Schematic diagram of the setup. Two superconductors are connected via a two normal metal sites. The Superconducting phases of the two superconductors are $\pm\phi_S/2$. Time dependent potential $V_A(t)$ and $V_B(t)$ are applied to sites A and B respectively.}~\label{fig-schem}
\end{figure}

We describe the system with the help of a tight binding model. Superconductors~(SCs) have $L_S$ sites and the NM  has just two sites. The superconducting phase of the left (right) SC is $\phi_S/2$ ($-\phi_S/2$).  The  superconductor on the left (right) is connected to the NM by hopping strength $w_L$ ($w_R$). Sinusoidal time dependent potentials $V_A(t), V_B(t)$ are applied to the NM sites $A, B$ respectively.
  $V_A(t)=V_0\cos{(\om t +\phi_{0})}$ and  $V_B(t)= V_0\cos{(\om t +\phi_{0}+\delta\phi_{V})}$. 
 The oscillating potentials are switched on at $t=0$. 
 The Hamiltonian for the system is  $H=H_0+H_1(t)\cdot \Theta(t)$, where $\Theta(t)=1$ for $t\ge 0$ and zero at other times and 
 \bea  
 H_0 &=& H_L + H_{LN} + H_N + H_{NR} + H_R, \nn \\
 H_L &=& -w\sum_{n=-1}^{-L_S+1}(c^{\dg}_{n-1}\tau_zc_{n}+{\rm h.c.})
 +\sum_{n=-1}^{-L_S}c^{\dg}_{n}\big[-\mu \tau_z\nn\\
 &&+\De\cos{(\f{\phi_S}{2})}\tau_x+\De\sin{(\f{\phi_S}{2})}\tau_y \big]c_{n}, \nn \\
H_R&=&-w\sum_{n=1}^{L_S-1}(c^{\dg}_{n+1}\tau_zc_{n}+{\rm h.c.}) 
+\sum_{n=1}^{L_S}c^{\dg}_{n}\big[-\mu \tau_z\nn\\ 
&&+\De\cos{(\f{\phi_S}{2})}\tau_x-\De\sin{(\f{\phi_S}{2})}\tau_y \big]c_{n}, \nn \\
 H_N&=& -\mu_0 c^{\dg}_{A}\tau_zc_{A} -\mu_0 c^{\dg}_{B}\tau_zc_{B}
            -w' (c^{\dg}_{A}\tau_zc_{B}+{\rm h.c.})  , \nn\\
H_{LN}&=&-w_L(c^{\dg}_{-1}\tau_zc_{A}+{\rm h.c.}),\nn\\
H_{NR}&=&-w_R(c^{\dg}_{1}\tau_zc_{B}+{\rm h.c.}),\nn\\
H_1(t)&=&V_A(t) c^{\dg}_{A}\tau_zc_{A}+V_B(t) c^{\dg}_{B}\tau_zc_{B}.~\label{eq:H-2site}
 \eea
 $c_n=[c_{n,\ua},-c_{n,\da},c^{\dg}_{n,\da},c^{\dg}_{n,\ua}]^T$, where $c^{\dg}_{n,\si}$ 
 creates an electron at site $n$ with spin $\si$ and $\tau_{x,y,z}$ are the 
 Pauli matrices that act on the particle-hole sector. Here, $w$ is the hopping amplitude, $\mu$ is the chemical potential, $\De$ is the magnitude of pairing in the superconducting lattices, $\phi_S$ is the superconducting phase difference, $L_S$ is the number of sites in each of the finite superconducting lattices, $w_L$ ($w_R$) is the hopping strength for the bond that connects the left (right) superconductor to the central NM region,  $\mu_0$ is the chemical potential in the NM region, $w'$ is the hopping strength on the bond that connects two sites in the NM region,  $V_A(t)$ and $V_B(t)$ are the time dependent onsite potentials on the two sites of the NM region.  Fig.~\ref{fig-schem} shows a schematic picture of the setup. Inversion operator  takes $n\to-n$, $A\to B$, $B\to A$. Time reversal operator  is $\kappa\si_y$, where $\kappa$ does complex conjugation and $\si_y$ is Pauli spin matrix acting on the spin space. Under time reversal, the superconducting phase $\phi_S$ changes to $-\phi_S$. It has to be noted that when the system is said to be invariant under time reversal or inversion, it refers to the invariance of the system for the case $\phi_S=0$. The system described by $H_0$ is invariant under inversion when $w_L=w_R$. 

 First, let us consider the time independent part of the Hamiltonian. A nonzero phase difference $\phi_S$ in drives a Josephson current $I_J$ at the junction. The Josephson current is the sum of  currents carried   by all eigenstates of $H_0$ with eigenenergies below zero. Let $\{|u_i\rangle,E_i,~i=1,2,..,N\}$ where $N=8(L_S+1)$  be the  eigenstates and eigenenergies of $H_0$ such that $E_i\le E_j$ for all $i<j$. The current operator at the bond $(A,B)$ is 
  \beq \hat J = \f{-iew'}{\hbar}(c^{\dg}_{A}c_{B}-c^{\dg}_{B}c_{A}), 
 \label{eq:curr-op}\eeq
where $e$ is the electron charge.
The Josephson current in the system described by $H_0$  is $I_J = \sum_{i=1}^{N/2} \langle u_i|\hat J|u_i\rangle$. The states $|u_i\ra$ can be calculated numerically exactly. 
 
 The Hamiltonian $H=H_0$ for $t<0$.  At time $t=0$, the system described by Hamiltonian $H$ moves away from the equilibrium ground state of $H_0$. The current at the junction for $t>0$ is aperiodic, despite the fact that the Hamiltonian is periodic in time. However, the current is a sum of two parts: one periodic in time and another aperiodic.  The aperiodic part vanishes when time averaged over the range $[0,\infty)$ and the periodic part survives averaging~\cite{soori10}. Such a long time averaged current $I_{av}$ is of central interest in this work, and we sketch the method to calculate $I_{av}$. Since the current $I_{av}$ draws contribution from periodic in time part only and this has a period $T$, the long time averaged current $I_{av}$ can be calculated by averaging the current over the time interval $[0,T]$.  If $T=2\pi/\om$ is the period of the time dependent part of the Hamiltonian, the interval $[0,T]$  is  sliced into $M$ equal intervals of size $dt=T/M$. The time dependent Hamiltonian is approximated by a constant Hamiltonian within  each of these smaller intervals. If $t_k$ is at the middle of the $k$-th interval, the operator that relates the state of the system at $t=T$ to the state of the system at time $t=0$ is given by 
 \beq U(T,0) = {\cal T} \prod_{k=1}^M {\rm exp}[-i H(t_k)dt], \label{eq:UT} \eeq
 where ${\cal T}$ is time ordering operator.   The eigenstates of $U(T,0)$: $|v_j\rangle$ have  eigenvalues $e^{i\th_j}$ and are termed Floquet states. The operator $U(T,0)$ and their eigenstates $|v_j\ra$ can be calculated exactly numerically since $H(t_k)$'s are known.
 For non-degenerate set of eigenstates within a spin subspace,  the  time averaged current $I_{av}$ can be expressed as 
\bea I_{av} &=& \sum_{i=1}^{N/2} \sum_{j=1}^N |c_{i,j}|^2 (J_T)_{jj},
{\rm ~~where~~} c_{i,j} =  \langle u_i|v_j\rangle \nn \\ 
&{\rm and} & ~~(J_T)_{jj} = \f{1}{T}\sum_{k=1}^M\langle v_j|U^{\dg}(t_k,
0)|\hat J|U(t_k,0)|v_j\rangle dt.~~~~ \label{eq:curr} \eea
Eq.~\eqref{eq:curr} points us to the fact that the Floquet states $|v_j\rangle$ carry the current. The overlap $c_{i,j}$ of the  initially occupied states $|u_i\rangle$ with the Floquet eigenstate $|v_j\ra$ dictates the weight of each Floquet state to the long time averaged current. In eq.~\eqref{eq:curr}, $\langle v_j|U^{\dg}(t_k,
0)|\hat J|U(t_k,0)|v_j\rangle = J_p(t_k)$ is the current carried by the Floquet state $|v_j\ra$ in the time interval of size $dt$ centered at $t=t_k$.  The current $I_{av}$ can be calculated numerically exactly.  The Floquet states are nonequilibrium states, and hence the diode effect exhibited by $I_{av}$ in the driven system is a nonequilibrium effect.

\section{Results and Analysis}~\label{sec-res}

Certain effects of driving in Josephson junctions have been studied in detail in a closely related model~\cite{soori20pump}. Our focus in this work will be to study Josephson diode effect by analyzing the current phase relation, where current refers to the long time averaged current $I_{av}$. 
\begin{figure}[htb]
 \includegraphics[width=8.6cm]{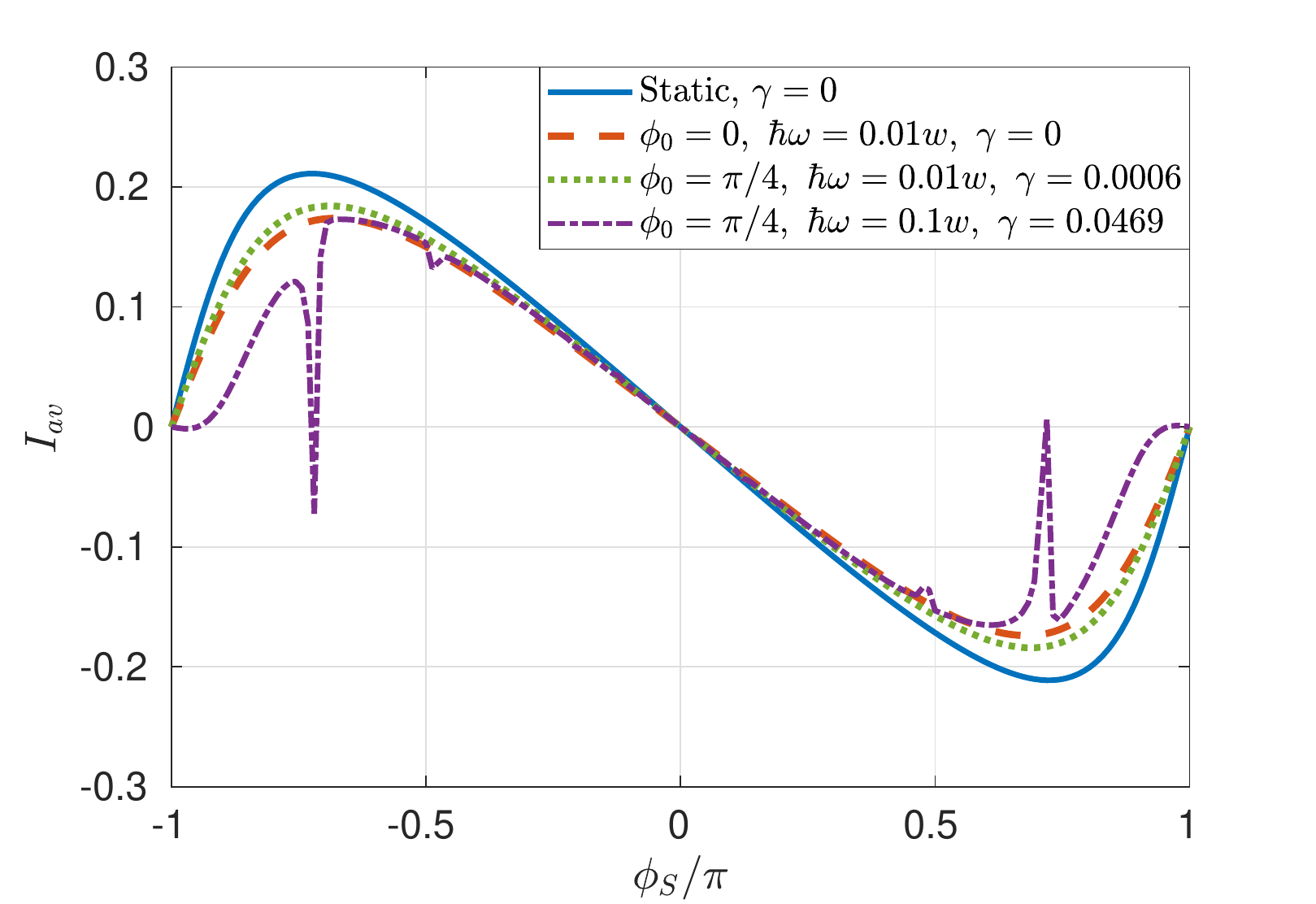}
 \caption{Current phase relation for a driven Josephson junction.  Current is in units of $ew/\hbar$. Parameters: $L_S=5$, $\De=0.5w$, $w'=0.3w$, $w_L=0.9w$, $w_R=0.5w$, $V_0=0.1w$, $\mu=0.01w$, $\mu_0=0$ and $M=150$. The solid blue line is the Josephson current for the undriven junction. Values of $\phi_0$, $\om$ and obtained $\ga$ are shown in the legend for each curve. }~\label{fig:cpr}
\end{figure}
Josephson critical current in the  forward (backward) direction $I_{c+}$ ($I_{c-}$) is  defined as the maximum (minimum) value of the current $I_{av}$ in the current phase relation. Typically, $I_{c+}>0$ and $I_{c-}<0$. Diode effect is characterized by $I_{c+}\neq -I_{c-}$ and is quantified by diode effect coefficient, defined as $\ga=2(I_{c+}+I_{c-})/(I_{c+}-I_{c-})$. A nonzero value of $\gamma$ indicates the diode effect.  The system does not exhibit diode effect if it is invariant under  inversion or time-reversal. 
\subsection{$w_L\neq w_R$}
In this subsection, we study diode effect by breaking inversion symmetry of $H_0$ by choosing $w_L\neq w_R$. 
We investigate the current phase relation for $\de\phi_V=0$ -the case when there is no pumping in the absence of a superconducting phase difference between the two superconductors. The dispersion in the superconductors is $E=\pm\sqrt{(2w\cos{k}+\mu)^2+\De^2}$, and the Josephson current is carried by subgap states that decay into the superconductor. For a state with $E\sim 0$, the decay length is close to $4$ for $\De=0.5w$, $\mu=0.01w$. Hence, a superconductor with $L_S>4$ sites mimics a sufficiently long superconductor and we choose $L_S=5$, $\De=0.5w$, $\mu=0.01w$. Current phase relations for $L_S=5$, $\De=0.5w$, $w'=0.3w$, $w_L=0.9w$, $w_R=0.5w$, $V_0=0.1w$, $\mu=0.01w$, $\mu_0=0$ and $M=150$ are shown in Fig.~\ref{fig:cpr}. The values of  $\phi_0$ and $\om$ for each curve are indicated in the legend. $\mu$ is chosen to be $0.01w$ to break the degeneracy of eigenstates. We find that for $\phi_0=0,\pi$ the diode effect is absent even in inversion asymmetric system since the system is invariant under time-reversal. For $\phi_0=0,\pi$, we find that the long time averaged  current $I_{av}(\phi_S)$ at a phase difference $\phi_S$  satisfies $I_{av}(\phi_S)=-I_{av}(-\phi_S)$.
In Fig.~\ref{fig:cpr}, the diode effect coefficient $\ga$ is indicated for each curve. This explains why the diode effect is absent for  for $\phi_0=0,\pi$. But for the choice $\phi_0=\pi/4$, the system exhibits diode effect.

\begin{figure}[htb]
 \includegraphics[width=4.211cm]{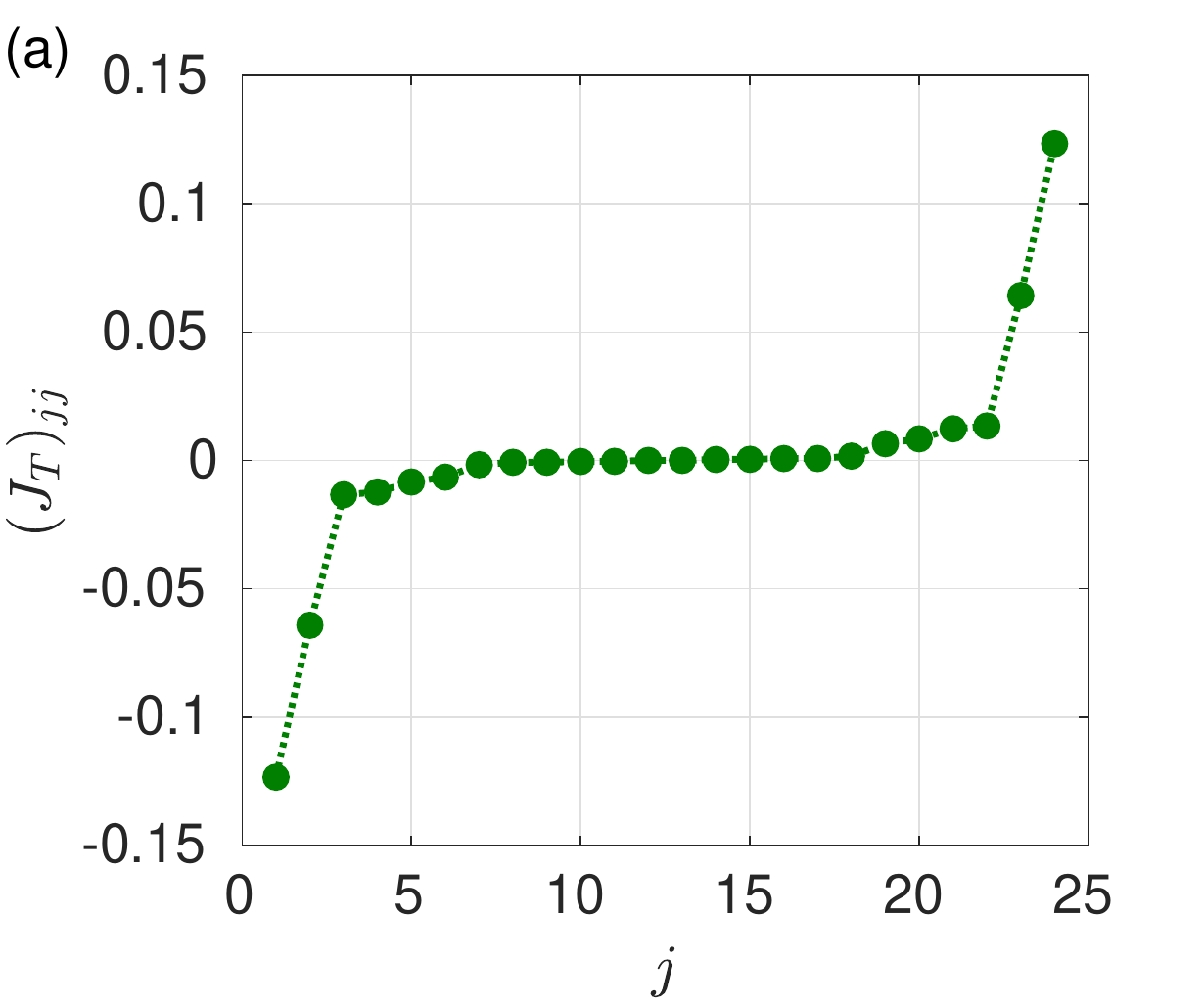}
  \includegraphics[width=4.21cm]{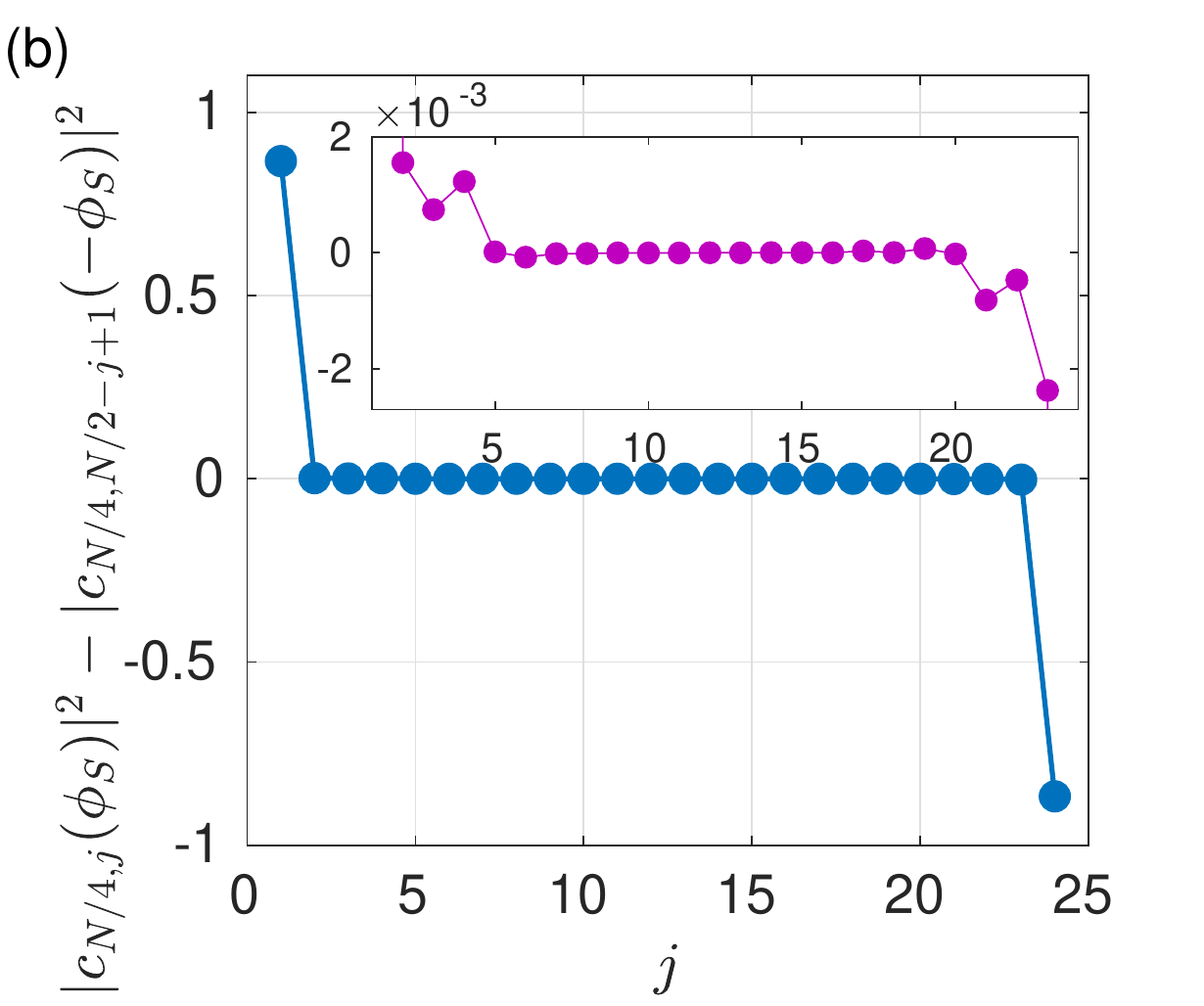}
 \caption{(a) Current $(J_T)_{jj}$ carried by the Floquet state labelled by $j$ sorted in ascending order of the current for superconducting phase differences $\pm\phi_S$, where $\phi_S=0.7\pi$. (b) The difference in the weights: $|c_{N/4,j}(\phi_S)|^2-|c_{N/4,N/2-j+1}(-\phi_S)|^2$ of Floquet states labelled by  $j$. The inset in (b) is zoomed $\hbar\om=0.1w$, $\phi_0=\pi/4$. Other parameters  are the same as in Fig.~\ref{fig:cpr}.}~\label{fig:expl-diode}
\end{figure}

\begin{figure}[htb]
 \includegraphics[width=4.2cm]{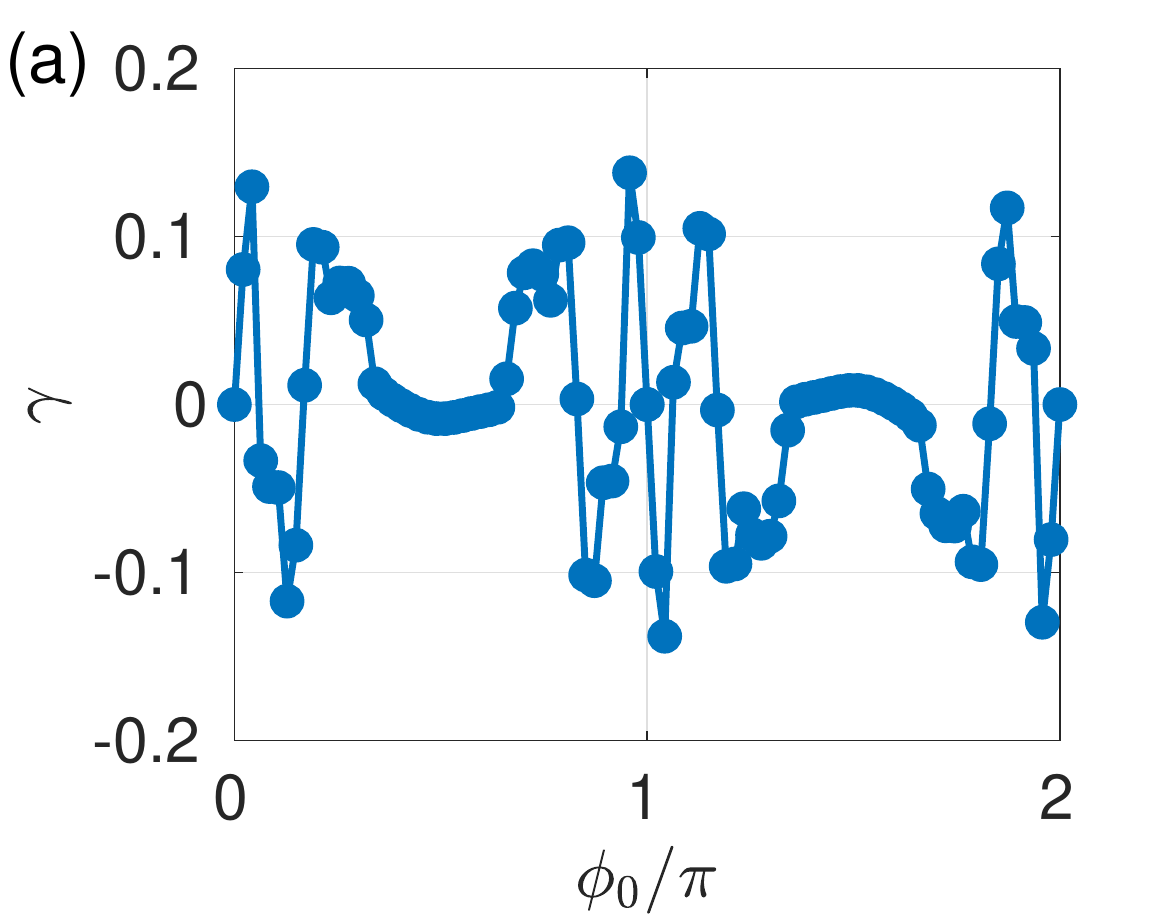}
 \includegraphics[width=4.201cm]{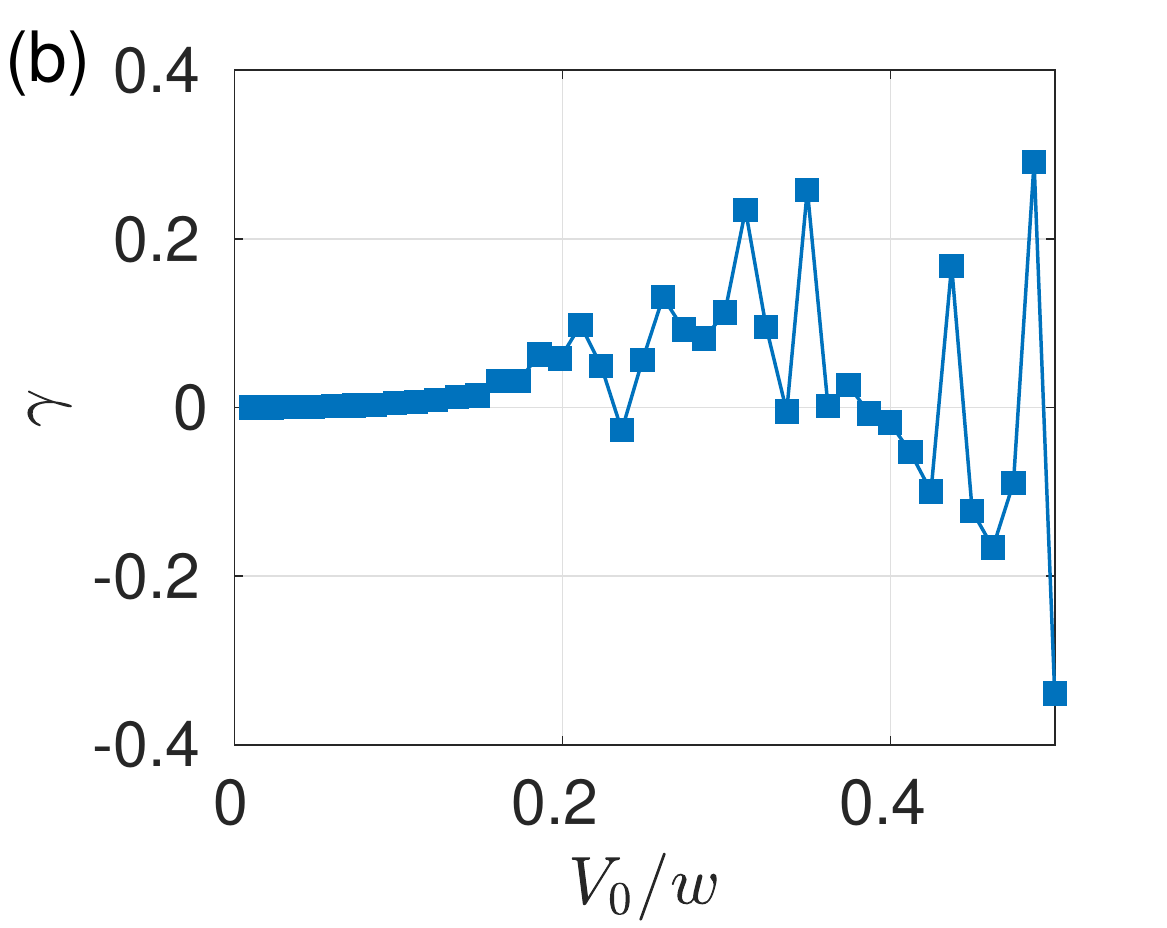}
 \caption{The dependence of diode effect coefficient on (a) $\phi_0$ for $V_0=0.25w$ and (b) $V_0$  for $\phi_0=\pi/4$. $\hbar\om=0.05w$ and other parameters are the same as in Fig.~\ref{fig:cpr}. }\label{fig:phiV0}
\end{figure}

The net Josephson current is carried by the Floquet states. 
To explain the results obtained, we analyze the current carried by a Floquet state at different times in a time period for two values of the superconducting phase differences: $\pm\phi_S$. For the case of $\phi_0=0$, the periodic part of the  current $J_p(t,\phi_S)$  at time $t$ for a superconducting phase difference $\phi_S$ satisfies $J_p(t,\phi_S)=-J_p(T-t,-\phi_S)$. The reason for this is that the ground state of the undriven system at superconducting phase differences $\phi_S$ and $-\phi_S$ are related by  time reversal operation. Time evolution over a time $t$ and $T-t$ are effectively time evolutions in opposite directions of time by the same duration since $U(T-t)|v_j\ra=e^{i\th_j}U(-t)|v_j\ra$, under the Hamiltonian $H_0+H_1(t)$. Further, the Hamiltonians at times $t$ and $-t$ are exactly the same. However, for the case $\phi_0=\pi/4$, the Hamiltonians at times $t$ and $-t$ are not the same.  In this case, the current $J_p(t)$ at time $t$ for $\phi_S$ is not equal to the magnitude of current for $-\phi_S$ at any other time in the time interval $[0,T]$. This makes  the time averaged currents for $\phi_S$ and $-\phi_S$ have different magnitudes and leads to diode effect. There is another way to understand the diode effect. The ordered (in ascending order) sets of average currents carried by the Floquet states for the SC phase differences $\phi_S$ and $-\phi_S$ are the same.  Further, the Floquet states come in pairs that carry time averaged currents which are equal in magnitude and opposite in sign. Fig.~\ref{fig:expl-diode}(a) shows the time averaged currents carried by the Floquet states for $\phi_S=\pm0.7\pi$. For $\phi_0=0,\pi$, the set of weights $|c_{i,j}(\phi_S)|^2$ for the choice of SC phase differences $\phi_S$ satisfies: $|c_{i,j}(\phi_S)|^2=|c_{i,N/2-j+1}(-\phi_S)|^2$. Here, we make use of the fact that the system is spin degenerate - up spin electron pairs with down spin hole and down spin electron pairs with up spin hole. So, we restrict the Hamiltonian to only the sector which has up spin electrons and down spin holes making the Hamiltonian an $N/2\times N/2$ matrix. Hence, the time averaged currents for phase differences $\phi_S$ and $-\phi_S$ are equal in magnitude and opposite in sign for $\phi_0=0,\pi$.  But, for $\phi\neq 0,\pi$, the set of weights $|c_{i,j}(\phi_S)|^2$ do not satisfy $|c_{i,j}(\phi_S)|^2=|c_{i,N/2-j+1}(-\phi_S)|^2$. This leads to the time averaged current $I_{av}(\phi_S)$ not being an odd function of $\phi_S$, finally resulting in diode effect. Fig.~\ref{fig:expl-diode}(b) shows the weights $|c_{N/4,j}|^2$ for $\phi_S=\pm0.7\pi$ and it is evident from the figure that weights do not satisfy: $|c_{i,j}(\phi_S)|^2=|c_{i,N/2-j+1}(-\phi_S)|^2$. The time averaged currents carried by the initial state $|u_{N/4}\ra$ for $\phi_S=0.7\pi, -0.7\pi$ are respectively $-0.0567ew/\hbar, 0.083ew/\hbar$. Fig.~\ref{fig:phiV0}(a) shows the dependence of the diode effect coefficient on $\phi_0$. We can see that the diode effect is absent for $\phi_0=0,\pi$.

Next, we examine the dependence of the diode effect coefficient $\ga$ on $V_0$ for $\phi_0=\pi/4$ keeping other parameters the same. The diode effect coefficient is plotted as a function of $V_0$ in Fig.~\ref{fig:phiV0}(b). The diode effect coefficient grows in magnitude as $V_0$ is gradually increased from $0$ and then oscillates with a large magnitude, even changing its sign. This is because, the minimum value of the difference between the energy levels of the undriven system close to zero energy $2E_b$ is $0.1038w$ for the chosen value of  parameters. When $V_0$ is larger than this value, the time dependent potential mixes the energy levels significantly and results in a large magnitude of the diode effect coefficient. As $V_0$ increases further from $2E_b$, the mixing of the energy levels of the undriven system is affected substantially, leading to oscillation in the value of the diode effect coefficient.

\begin{figure}[htb]
\includegraphics[width=8cm]{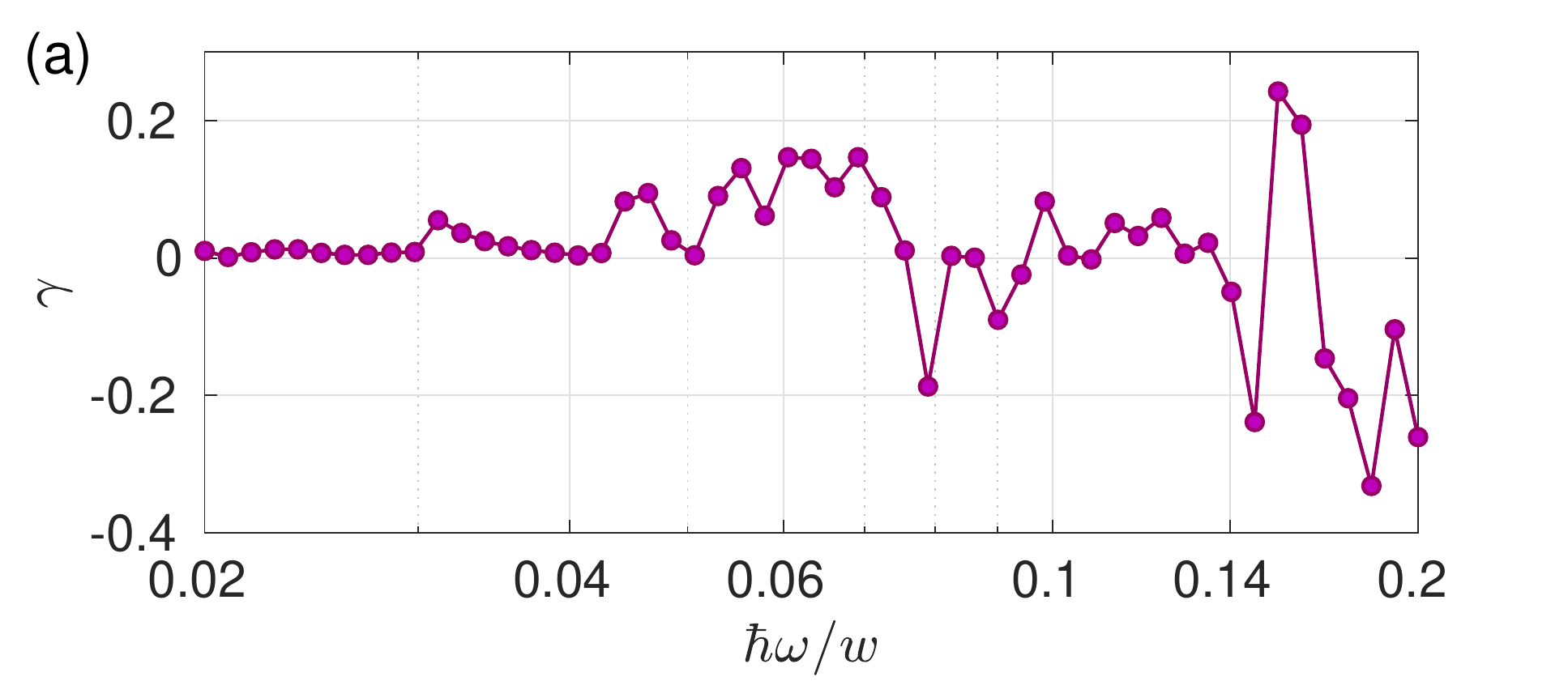}
 \includegraphics[width=8cm]{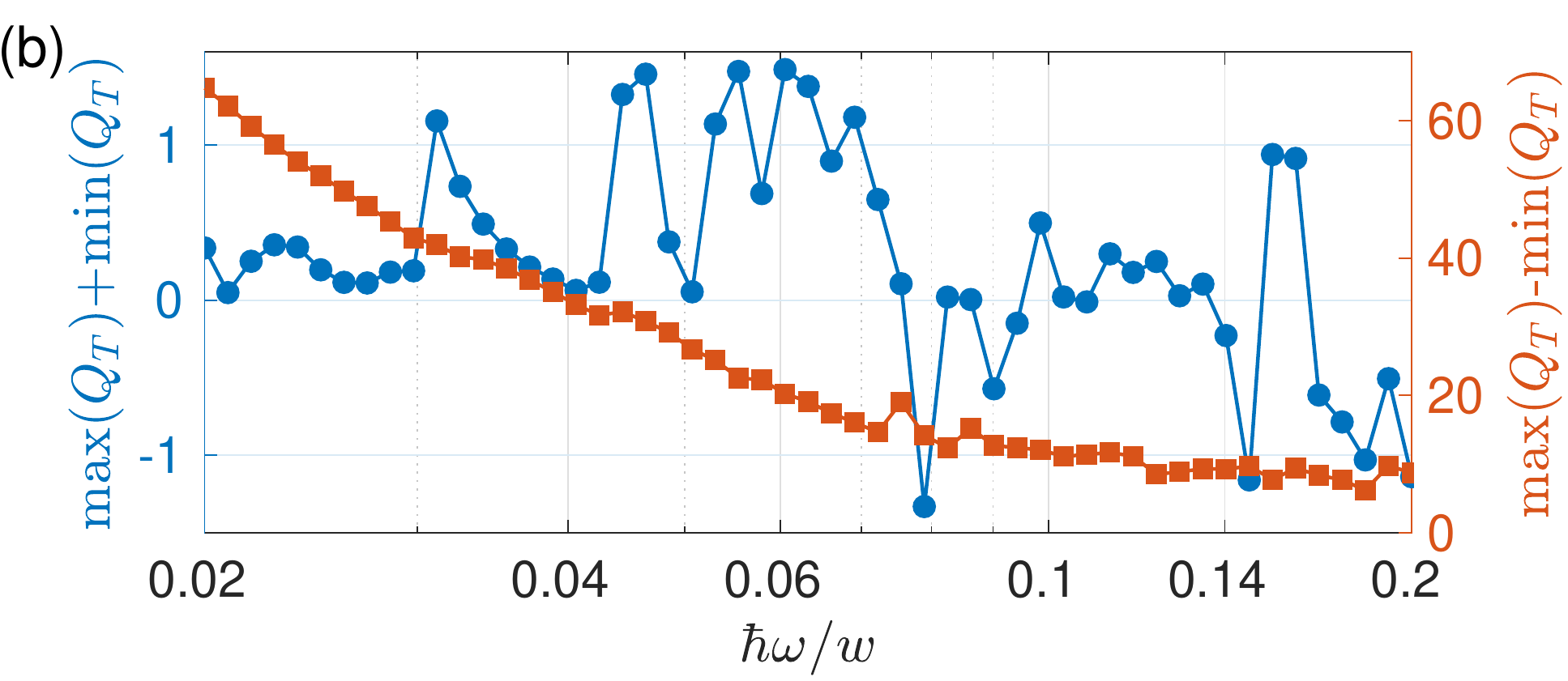}
\caption{(a) Diode effect coefficient as a function of the frequency $\om$.  (b) The sum of (left ordinate) and the difference between (right ordinate) ${\rm max}(Q_T)$ and ${\rm min}(Q_T)$ plotted versus the frequency $\om$. $V_0=0.25w$ and $\phi_0=\pi/4$ and other parameters are the same as in Fig.~\ref{fig:cpr}.}\label{fig:diode-om}
\end{figure}

Now, we study - `how does the change in frequency of the periodic potential affect diode effect?' In the numerical calculation, the time interval $[0,T]$ over which the current is averaged  is sliced into $M$ small intervals and in each of these intervals of duration $dt$, the time dependent potentials are treated as constants. Hence, while studying the dependence of the diode effect coefficient as a function of $\om$, we keep the duration $dt$ almost the same as the frequency of the periodic potential is varied. We show the dependence of diode effect coefficient on frequency in Fig.~\ref{fig:diode-om}(a) for $V_0=0.25w$, keeping other parameters the same. We choose $M=[5T\De/\hbar]$, where $[x]$ corresponds to the largest integer less that $x$.   
It can be seen that the diode effect coefficient approaches zero in the adiabatic limit. In the adiabatic limit, charge transferred across the two electrodes over one time period $Q_T$ is calculated. To comprehend the dependence of the diode effect coefficient in the adiabatic limit, the sum $[{\rm max}(Q_T)+{\rm min}(Q_T)]$ and difference $[{\rm max}(Q_T)-{\rm min}(Q_T)]$ are plotted as functions of the frequency in Fig.~\ref{fig:diode-om}(b). The diode effect coefficient is related to $Q_T$ by $\ga=2[{\rm max}(Q_T)+{\rm min}(Q_T)]/[{\rm max}(Q_T)-{\rm min}(Q_T)]$. It can be seen that  though the maximum charge transferred in one time period between the phase biased superconductors diverges as $\om\to0$, the difference between the magnitudes of the maximum charge transferred in the forward and backward directions remains almost the same in magnitude as $\om\to0$ making the diode effect coefficient approach zero in the adiabatic limit.

\begin{figure}[htb]
 \includegraphics[width=8cm]{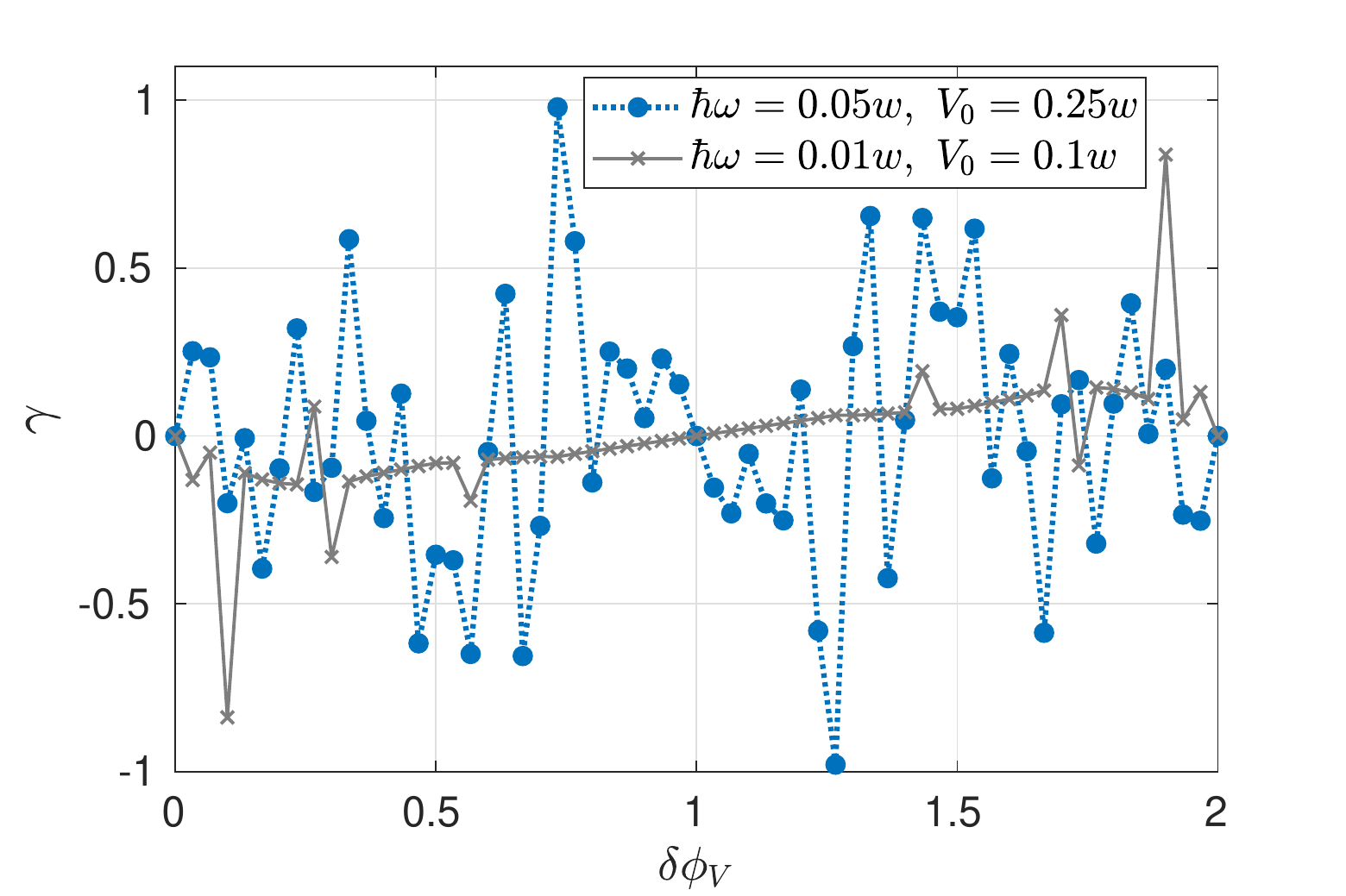}
 \caption{Diode effect coefficient versus $\de\phi_V$ - the difference in phases of the time dependent onsite potentials for $\phi_0=0$, $\hbar\om=0.05w$, $V_0=0.25w$, $w_L=w_R=0.9$, $M=50$ and other parameters same as in Fig.~\ref{fig:cpr}.  }\label{fig:phiv}
\end{figure}

\subsection{$H_{1}(t)$ breaks inversion and time reversal symmetries}
In this subsection, we study the case in which the time-independent part of the Hamiltonian is inversion symmetric, but the inversion and time reversal symmetries are broken by the time dependent part of the Hamiltonian. In such a case, the diode effect is absent for $\de\phi_V=0, \pi$ since there is no preferential direction in which the current is favored. The ordered sets of time averaged currents carried by the Floquet states for the superconducting phases differences $\pm\phi_S$ are not the same in this case in contrast to the first method wherein $\de\phi_V=0$.   However, for $\de\phi_V\neq0,\pi$, a nonzero current is pumped even at $\phi_S=0$ and the time dependent potentials break inversion and time reversal symmetry. This gives a preferential direction for the flow of current, and the system exhibits  diode effect. In Fig.~\ref{fig:phiv}, the diode effect coefficient is plotted as a function of the difference $\de\phi_V$ in phases of the time-dependent potentials for $w_L=w_R=0.9w$, $\phi_0=0$, $\hbar\om=0.05w$ and $V_0=0.25w$. We see that the diode effect coefficient is nonzero for $\de\phi_V\neq 0,\pi$. Further, we find that the diode effect coefficient tends to zero in the adiabatic limit despite a finite amount of charge being transferred from one SC to another over one time period, which can be explained using figures similar to Fig.~\ref{fig:diode-om}. 

\begin{figure}[htb]
 \includegraphics[width=4.2cm]{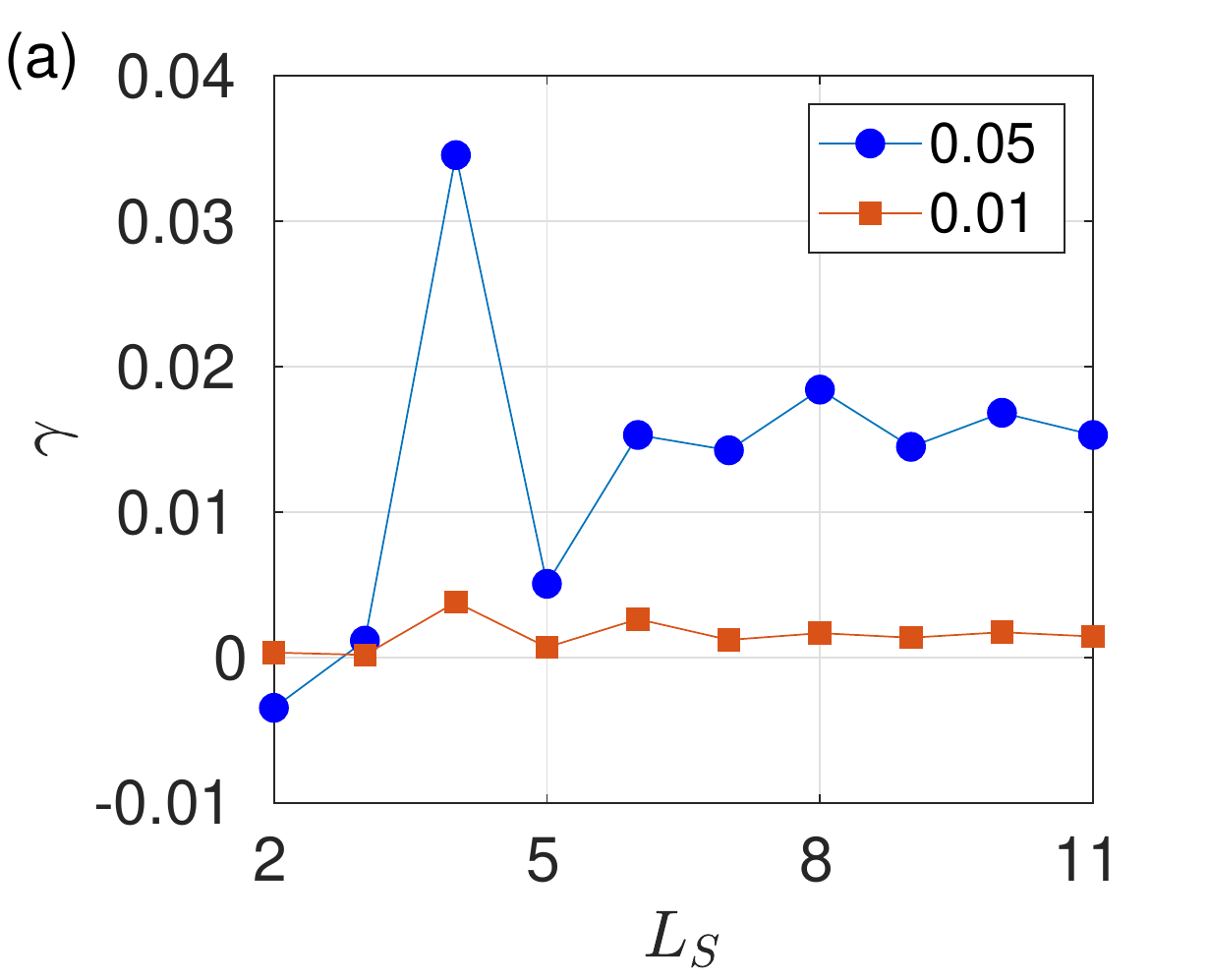}
 \includegraphics[width=4.2cm]{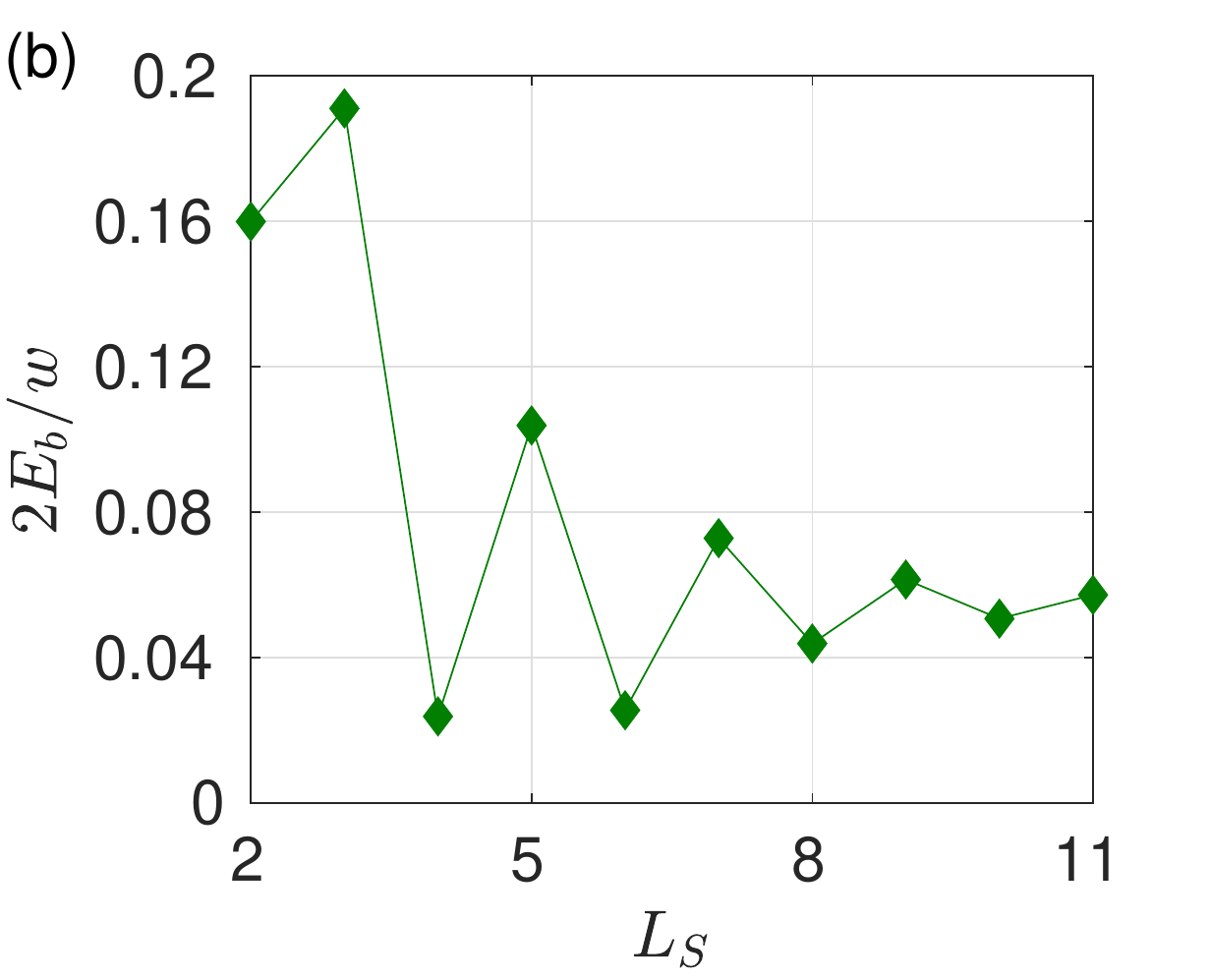}
 \caption{(a) Diode effect coefficient versus the length $L_S$ for values of parameters same as in Fig.~\ref{fig:cpr}. Driving frequency $\hbar\om/w$ for each curve is indicated in the legend. (b) Minimum value of the difference in the eigenenergies of $H_0$ as $\phi_S$ is varied in units of $w$ versus $L_S$. $\phi_0=\pi/4$ and other parameters are the same as in Fig.~\ref{fig:cpr}.}\label{fig:LS}
\end{figure}

\subsection{Dependence on $L_S$}
In our calculations so far, we have taken $L_S=5$. In this section, we analyze the dependence of the diode effect on $L_S$ -the size of the superconductor. For small driving frequencies, eigenstates $H_0$ close to zero energy participate in the dynamics. For energies within the superconducting gap, the wave function in the superconductor decays away from the junction and for zero energy state, the wave function diminishes by a factor $1/e$ over a length $\sim 4$ for the choice of parameters used in the  paper. In Fig.~\ref{fig:LS}(a), we plot the diode effect coefficient $\ga$ versus the length $L_S$ of the SC region. In Fig.~\ref{fig:LS}(b), the dependence of $2E_b$ - the minimum value of the difference between energy levels of $H_0$ closest to zero  as $\phi_S$ is varied is plotted versus the size of the superconductor. It is evident that the diode effect coefficient is peaked at $L_S=4$ and saturates as $L_S$ increases. This is because, for smaller values of $L_S$, the charge  driven from one SC to another  reflects back from the boundary of the second SC and returns back to the first SC. As the $L_S$ becomes larger, the driven charge relaxes in the second SC. Also,  $E_b$ depends on  $L_S$ and comparing subplots (a) and (b) in Fig.~\ref{fig:LS}, one can infer that for smaller values of the energy difference $2E_b$, driving results in higher value of $\gamma$. For $\hbar\om=0.05w$, the driving makes transition between energy levels of $H_0$ closest to zero possible, 
whereas for $\hbar\om=0.01w$, driving can help transition between energy levels of $H_0$ closest to zero only through a higher order processes~\cite{moskalets02}. The energy difference $2E_b$ for larger length $L_S$ is much closer to the driving frequency for $\hbar\om=0.05w$ and results in much higher values of $\gamma$. 

\section{Summary  and Conclusion}~\label{sec-con}
We have proposed a setup where Josephson diode effect can be realized by periodic driving of the NM region in an SNS junction. Time reversal and inversion symmetries need to be broken for the diode effect to show up. There are two ways in which diode effect can be accomplished in the system: one by breaking inversion symmetry in the undriven part of the Hamiltonian by choosing $w_L\neq w_R$ and another by breaking symmetries purely in the driven part of the Hamiltonian by choosing $\de\phi_V\neq0,\pi$. Naturally, a combination of the two ways wherein both $w_L\neq w_R$ and $\de\phi_V\neq0,\pi$ can also result in diode effect. We have studied the dependence of the diode effect coefficient on the amplitude of the driving potential $V_0$ and frequency $\om$. We find that though the charge may be pumped in the adiabatic limit, the diode effect vanishes in the adiabatic limit in both the schemes. The setup in this work that exhibits diode effect is a two terminal setup, in contrast to many four terminal setups~\cite{song98,ideue17,siva18,ando2020,legg22,moodera}. 

In a realistic setup, it is possible that the periodic driving heats up the system and the low temperature environment cools the system. This effect is not captured by the formalism here. Also, such a process takes much longer times compared to the time scale of driving and it is unlikely that it hinders the experimental realization of the proposed phenomenon. 

Recently,  Josephson junction with graphene in the middle driven with the help of microwave irradiation has been experimentally realised~\cite{park2022}. 
The ability to periodically modulate the potential in quantum dots connected to superconductors~\cite{Cleuziou2006,DeFranceschi2010,Wan2015} forms an important step in experimentally realizing the proposed setup. Purely metallic systems driven by periodic potentials that exhibit pumping have been realized experimentally~\cite{switkes99,Drexler2013}. Hence, the setup proposed is  within the reach of present technology.  Josephson junctions promise development of quantum computer~\cite{clarke2008,devoret13} and have influenced quantum technologies~\cite{Kleiner07,lee2020}.  Also, Josephson junctions have been proposed to be platforms in braiding of non-Abelian Majorana fermions~\cite{vanheck2012}. Hence, realization of nonequilibrium Josephson diodes proposed in this work will add to an assortment of quantum devices that will shape the future research.

While revising the manuscript, we became aware of ref.~\cite{paaske} where rectification and anomalous Josephson current have been studied in a setup similar to ours. 

\section*{ Acknowledgements} The author thanks Udit Khanna and Diptiman Sen for useful discussions. The author thanks Dhavala Suri for bringing his attention to the phenomenon of superconducting diode effect and discussions. The author acknowledges financial support by DST through DST-INSPIRE Faculty Award (Faculty Reg. No.~:~IFA17-PH190). 
\bibliography{refsde}

\end{document}